\documentclass[runningheads]{llncs}

\usepackage{graphicx}
\begin{document}
\title{ElectroAR: Distributed Electro-tactile Stimulation for Tactile Transfer}
%
%
\author{Jonathan Tirado\inst{1} \and
Vladislav Panov\inst{1} \and
Vibol Yem\inst{2} \and
Dzmitry Tsetserukou\inst{1} \and
Hiroyuki Kajimoto\inst{3}}
\authorrunning{J.Tirado, V.Panov et al.}
%
\institute{Skolkovo Institute of Science and Technology (Skoltech), Moscow 121205, Russia
\email{\{jonathan.tirado,vladislav.panov,D.tsetserukou \}@skoltech.ru}\\
\and
Tokyo Metropolitan University, Tokyo Japan
\email{yemvibol@tmu.ac.jp}
 \and
The University of Electro-Communications, 1-5-1 Chofugaoka, Chofu, Tokyo 182-8585, Japan 
\email{kajimoto@kaji-lab.jp}\\
}
\maketitle              
\begin{abstract}
We present ElectroAR, a visual and tactile sharing system for hand skills training. This system comprises a head-mounted display (HMD), two cameras, a tactile sensing glove, and an electro-tactile stimulation glove. The trainee wears the tactile sensing glove that gets pressure data from touching different objects. His/her movements are recorded by two cameras, which are located in front and top side of the workspace. In the remote site, the trainer wears the electro-tactile stimulation glove. This glove transforms the remotely collected pressure data to electro-tactile stimuli. Additionally, the trainer wears an HMD to see and guide the movements of the trainee. The key part of this project is to combine distributed tactile sensor and electro-tactile display to let the trainer understand what the trainee is doing. Results show our system supports a higher user's recognition performance.
\end{abstract}
\section{Introduction}
There are several tasks that incorporate hand-skill training, such as surgery, palpation, handwriting, etc. We are developing an environment where a skilled person (trainer), who actually works at a different place, can collaborate with a non-skilled person (trainee) in high precision activities. The trainer needs to feel as if he/she exists at the place and work there. The trainee can improve his/her performance with the trainer’s help. This can be regarded as one type of telexistence \cite{Susumu}, in which remote robot is replaced by trainee.

We especially focus on the situation that incorporates finger contact. This requires a tactile sensor on the trainee's side and tactile display on the trainer's side. The trainee handles real objects, and the tactile sensor-display pair enables the trainer to feel the same tactile experience as the trainee; thus, he/she can command or show what the trainee should do next.

For tactile sensors, a wide variety of these pads have been developed in the past for robotics and medical applications, using resistive, capacitive, piezoelectric, or optical elements. These pads have often been placed in gloves to monitor hand manipulation. While some of them are bulky and inevitably deteriorate the human haptic sense, recently, several researches are focused on reducing this problem by using thinner and more flexible force-sensing pads \cite{Beebe}. In this study, we use a similar tactile sensor array with high spatio-temporal resolution.  

For tactile display, there were also several researches on wearable tactile displays \cite{Choi}. They are simple, yet cannot present distributed tactile information that our sensor can detect. As we believe that distributed tactile information is important, especially when we recognize shapes, we need some way to present distributed tactile information to fingertips. There were also several works on pin-array type tactile display \cite{Kim kang}, \cite{Sarakoglou}. We employ electro-tactile display \cite{Saunders}, \cite{P.Bach}, since it is durable, light-weight, and easy to be made small and extends to several fingers. 

This paper is an initial report of our system, especially focuses on how well the shape information can be transmitted through our system.

\begin{figure}
\begin{center}
\includegraphics[width=0.92\textwidth]{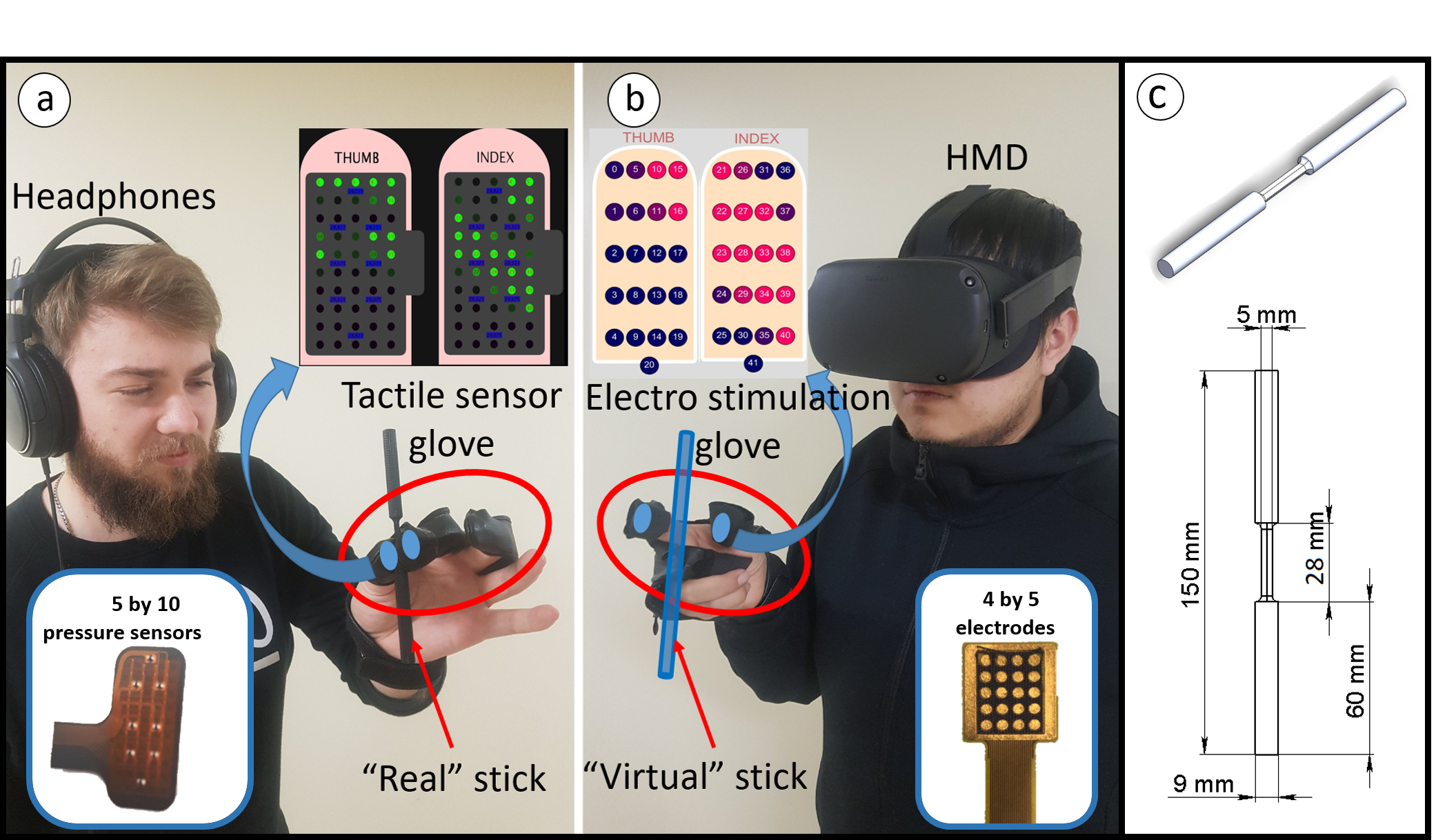}
\end{center}
\caption{ElectroAR. (a) Follower side. (b) Leader side. (c) Cylindrical stick with regular prismatic shape} \label{fig1}
\end{figure}
\section{System overview}
As shown in Fig.1, the system consists of three main components. On the follower (trainee's) side, the user wears a tactile sensing glove. The glove gets the pressure data from touching objects. 
The data of pressure sensors were spatially filtered by using equation (\ref{E:relation}), 

\begin{equation}\label{E:relation}
	p\textsc{\char13}_{i,j}=\frac{p_{i,j}+p_{i+1,j}+p_{i,j+1}+p_{i+1,j+1}}{4}
\end{equation}	
where \textit{$p$} is a pressure value and \textit{$p$}\textsc{\char13} is a filtered pressure value, i and j are order number on the axis of width and height of the sensor array \cite{Yem Kaji}.

The leader's glove transforms the filtered pressure data to electro-tactile stimuli at fingertips. They are linked not only with haptic feedback but also with visual and audio feedback. 
Visual feedback gives for the leader side full information of the movement on the follower side, and audio feedback provides for the follower side commands from the leader.

\subsection{Tactile sensor glove}
We are using a glove that contains three tactile sensor arrays \cite{Yem Kaji}. These sensor arrays are located on the three fingers of the right hand (thumb, index and middle). Fig.1 (a) shows the internal distribution of the pressure sensors in the array of 5 by 10 for each finger. The force range of sensing element was not accurately measured, but it can discriminate edge shapes by natural pressing force, as will be shown in the experiment section. The center-to-center distance between each sensing point is 2.0 \textit{$mm$}. 

\subsection{Electro-tactile glove}
Fig.1 (b) shows the glove of electro-tactile display for the leader user \cite{Yem Kaji}. The module controller was embedded inside the glove \cite{Kaji electro}. For each finger, the electro-tactile stimulator array has 4 by 5 points. The center-to-center distance between each point is 2.0 \textit{$mm$}. This module was used for tactile stimulation of thumb, index and middle finger. The pulse width is set to 100 \textit{$us$}.


\subsubsection{Random Modulator}
In order to adjust the intensity of the stimulus, a typical method is to express the intensity by a pulse frequency. However, in practice, the stimulator must communicate with the PC at fixed intervals (in our case at 120 \textit{$Hz$}). Therefore, although it is relatively easy to set the pulse frequency to, for example, 30 \textit{$Hz$}, 60 \textit{$Hz$}, or 120  \textit{$Hz$}, it is a little difficult to perform electrical stimulation of an arbitrary frequency.

Here, we propose a method to change the probability of stimulation as a substitute for setting pulse frequency. For each time interval (in our case 1/120 \textit{$second$}), the system gives the probability of stimulating each electrode. The higher the probability, the higher the average stimulus frequency. The algorithm is expressed as follows.

\begin{equation}\label{E:relation1}
   \textbf{if}\:\;\;rand\:() \leq  p\:\;\;\textbf{then}\:\;\; stimulate\:() 
\end{equation}

Where \textit{$rand \: ()$} is a uniformly distributed random variable from 0 to 1. If it is less than or equal to a value \textit{$p$}, the electrode is stimulated. Otherwise, it is not stimulated. The probability that the electrode is stimulated is hence \textit{$p$}. This calculation is performed for the electrode every cycle, resulting in an average stimulation cycle of \textit{$120*p$} \textit{$Hz$}.

The value \textit{$p$} represents the probability of stimulation, and a function representing the relationship between \textit{$p$} and the subjective stimulus intensity \textit{$S$} is required. In general, higher stimulus frequency gives stronger subjective stimulus, so this function is considered to be a monotonically increasing function.

\begin{equation}\label{E:srelation}
   S = F(p) 
\end{equation}

If \textit{$F$} is obtained, the inverse function can be used to determine how the stimulus probability \textit{$p$} should be set for the intensity \textit{$S$} to be expressed as follows.

\begin{equation}\label{E:prelation}
   p = F^{-1}(s) 
\end{equation}



\subsection{View sharing system}

Ideally, the view sharing system should be bi-directional. However, as the scope of this paper is to examine the ability of our tactile sensor-display pair, we used a simplified visual system only for the trainer.

As shown in Fig.1 (b), the trainer wears an HMD. At the remote side, two cameras are installed for having full view information for the trainer, both from the top and from the side. This information is presented in virtual screens which are located in front and the horizontal view. Although the view is not three-dimensional, it can provide sufficient information of the trainee's hand movement, and the trainer can mimic the movement while perceiving the tactile sensation by the electro-tactile display glove.

\section{Experiment}
\subsection{Preliminary Experiment : Random modulator's function}

The proposed random modulation method needs a function \textit{$F$}, which can represent the relationship between strength perception and the probability of stimulating each electrode. This preliminary experiment has the objective of collect data for fitting function \textit{$F$}. In the whole experiment, the base stimulation frequency was 120 \textit{$Hz$}. For example, if the probability is 1, the stimulation is done at 120 \textit{$pps$} (pulses per second). 

\subsubsection{Experimental Method}

The strength of stimulation was evaluated by the magnitude estimation method. First, the user's right index fingertip was put on the electrodes' array, and exposed to a pulsatory stimulation, provided by electrodes.
The user was asked to find a comfortable and recognizable level (absolute stimulation level), which was set as 100.

In the second part, we prepared six probability levels: 0.1, 0.2, 0.4, 0.6, 0.8, 1.0. There were five trials for each level, 30 trials in total in random order. Each trial was composed of an initial one-second impulse with the 100 intensity level, followed by a one-second randomly modulated stimulation with assigned probability. After each trial, the user must determine how lower or higher was the second stimulus presented. We recruited seven participants, five males and two females aged 21-27; all right-handed and all without previous training.  

\subsubsection{Result}
\begin{figure}
\begin{center}
\includegraphics[width=1.02\textwidth]{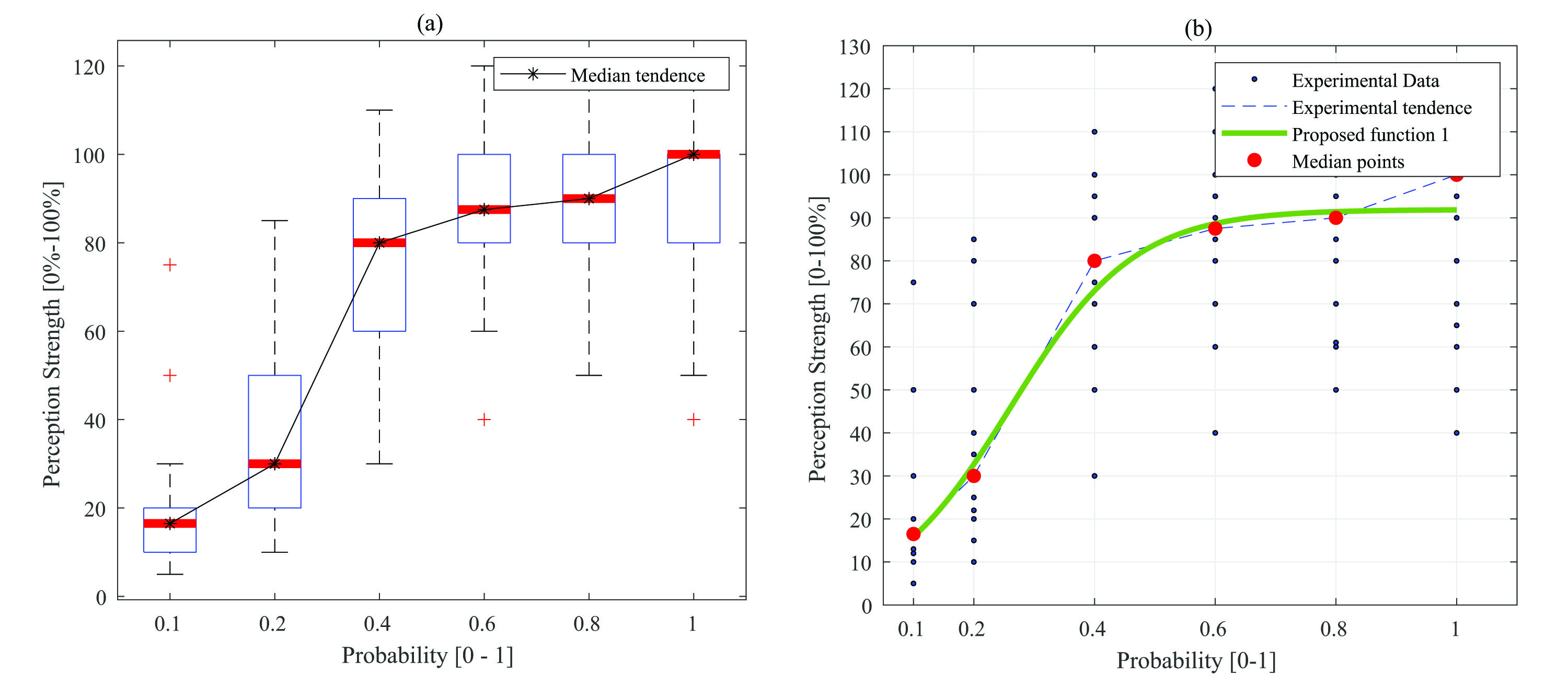}
\end{center}
\caption{Random Modulation. (a) Experimental results. The quantitative relation between cumulative probability distribution and the strength perception percentage estimated for the volunteers. (b) Sigmoid function regression. Experimental data were fitted to sigmoid function by logistic regression}\label{bs1}
\end{figure}

The result in Fig.\ref{bs1} (a) shows a sigmoid function tendency. Thus, the data were fitted using Matlab, as shown in Fig.\ref{bs1} (b).  

 Once we know the function, we calculated the inverse function that determines the stimulation probability from desired strength, which is the function \textit{$F^{-1}$}, described in the equation (\ref{eq:3}), where \textit{$a,b$} and \textit{$k$} are coefficients of the sigmoid function, \textit{$p$} is the probability of the electrode being stimulated and \textit{$S$} is the subjective stimulus intensity.
 
\begin{equation}\label{eq:3}
p=\frac{a-log(\frac{k}{S}-1)}{b}
\end{equation}

\subsection{Experiment 1: Static shapes recognition}
The following two experiments try to validate that our system is capable of transmitting tactile information necessary for tactile skill transfer. In many haptic related tasks, we typically use a pen-type device that we pinch by our index finger and thumb. These can be a scalpel, a driver, a tweezer, or a pencil. In such situations, we identify the orientation of the device with tactile sense. 

Our series of experiments try to reproduce part of these situations. Experiment 1 was carried out to assess the electro-tactile display's capacity for presenting bar-shape in different orientations.     

\subsubsection{Experimental Method}

Four patterns, which are line with inclinations of 0, 45, 90 and 135 \textit{$degrees$} were presented on the right index finger. The experiment was divided into three steps.
The first step was to identify a suitable stimulus level. The second step was the training phase, in which each pattern was presented twice to the volunteers. 

After a two minutes break, the evaluation stage was performed. They were asked to try randomly chosen pattern and chose from the four candidates. The recognition time was also recorded.
We recruited ten volunteers, nine males and one female, aged 21-27; all right-handed. There were seven trials per pattern, 28 in total. 

\subsubsection{Result}

\begin{figure*}[h]
\begin{center}
\includegraphics[width=1.02\textwidth]{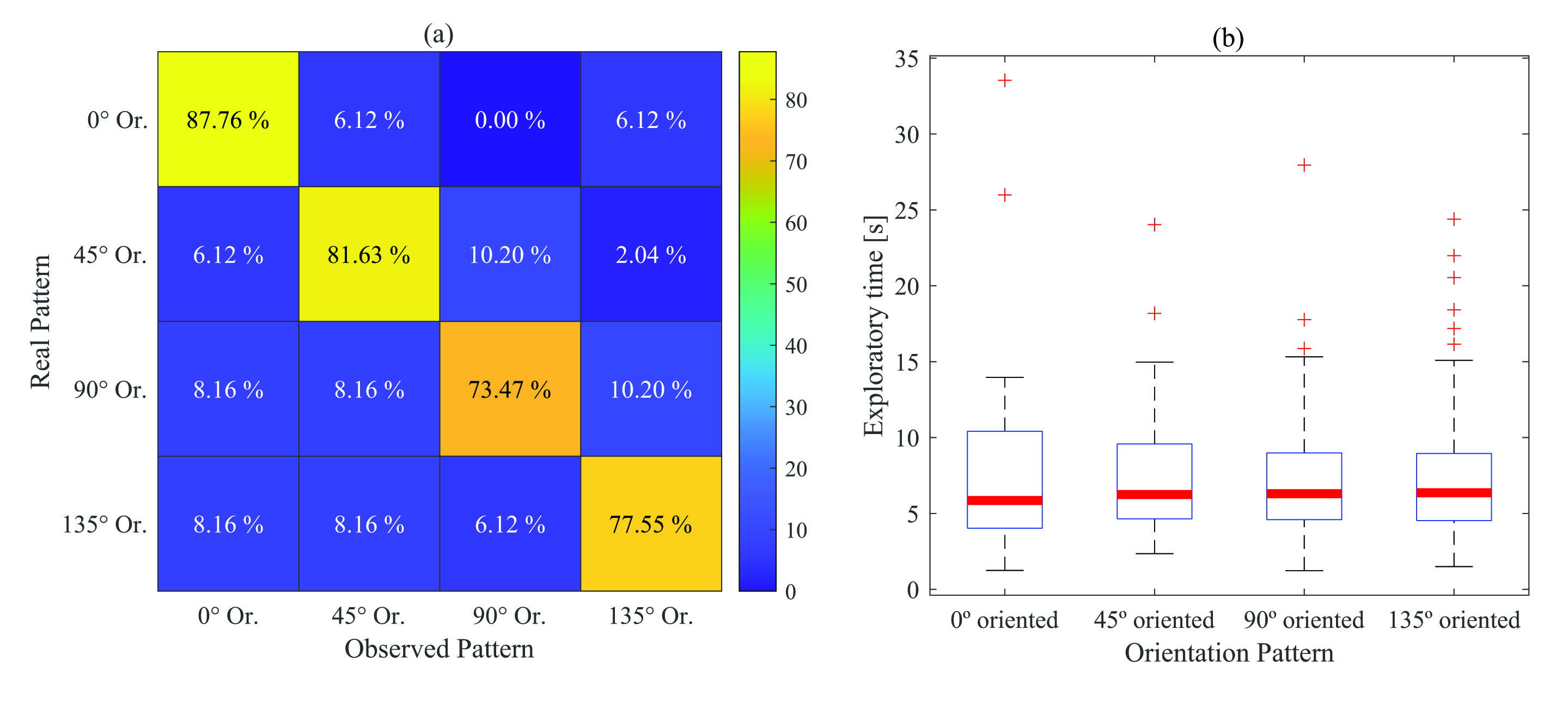}
\end{center}
\caption{Experiment 1. (a) Confusion Matrix Pattern recognition rate. (b) Exploratory time comparative evaluation}\label{bs2}
\end{figure*}

Fig.\ref{bs2} (a) shows a numerical comparison of the effective recognition level for each proposed pattern. The four patterns have a similar range of recognition, being the 90 \textit{degrees} pattern pointed the lowest (73\% accuracy) and the 0 \textit{degrees} pattern the highest (87\% accuracy). The result also indicates that the 90 \textit{degrees} pattern is often  confused with the 135 \textit{degrees} (10\% error), and in the same way the 45 \textit{degrees} is confused with 90 \textit{degrees} pattern (10\% error). 

Fig.\ref{bs2} (b) shows that the recognition time for the majority of the volunteers ranges between 4 and 10 \textit{seconds} for all of the patterns. The median time is close to 6 \textit{seconds}. 

\subsection{Experiment 2: Dynamic pattern perception}
Experiment 2 was carried out to assess our system’s capacity to convey dynamic tactile information. As mentioned before, we focused on the situation of handling a bar-shaped device. We confirmed if we can identify different “devices” that we handle with our index finger and thumb.

\subsubsection{Data set acquisition}

Four cylindrical sticks with regular prismatic shape in their middle section were designed for the experiment (Fig.1 (c)). The total length of each stick is 150 \textit{mm}, and 28 \textit{mm} for their middle section. Every prism has a different cross-section: circle, triangle, square, and hexagon. The radius of the sticks was 9 \textit{mm} and the circumradius of the prisms 5 \textit{mm}. This special design visually covers the middle section for avoiding the possibility of answering only by observation. On this way, we provide only the motion of the hand as visual feedback.
 
Using the tactile sensing glove, one of the authors grasped the stick in a 90 \textit{degrees} orientation, and he slowly scrolled the bar back and force between two fingers, repeating for ten times. The pressure patterns were recorded, and the video was taken by two cameras that we described in the previous section. 

\subsubsection{Experimental Method}
In the main experiment, the recorded videos were replayed so that the user can mimic the hand motion. Simultaneously, the tactile feedback was delivered to two fingertips (right index finger and thumb) using the recorded pressure patterns.
 
A set of twenty randomly ordered samples was presented, and the user must associate this visual and tactile sensation with one of the previously indicated shapes. Visual feedback was provided to show the motion of the hand, but at the same time, the shape of the prism was visually hidden. The recognition time was also recorded. 
 
We recruited eight participants, six males and two females aged 21-27; all right-handed and all without previous training.  
 
 \subsubsection{Result}
 
\begin{figure*}[h]
\begin{center}
\includegraphics[width=1.02\textwidth]{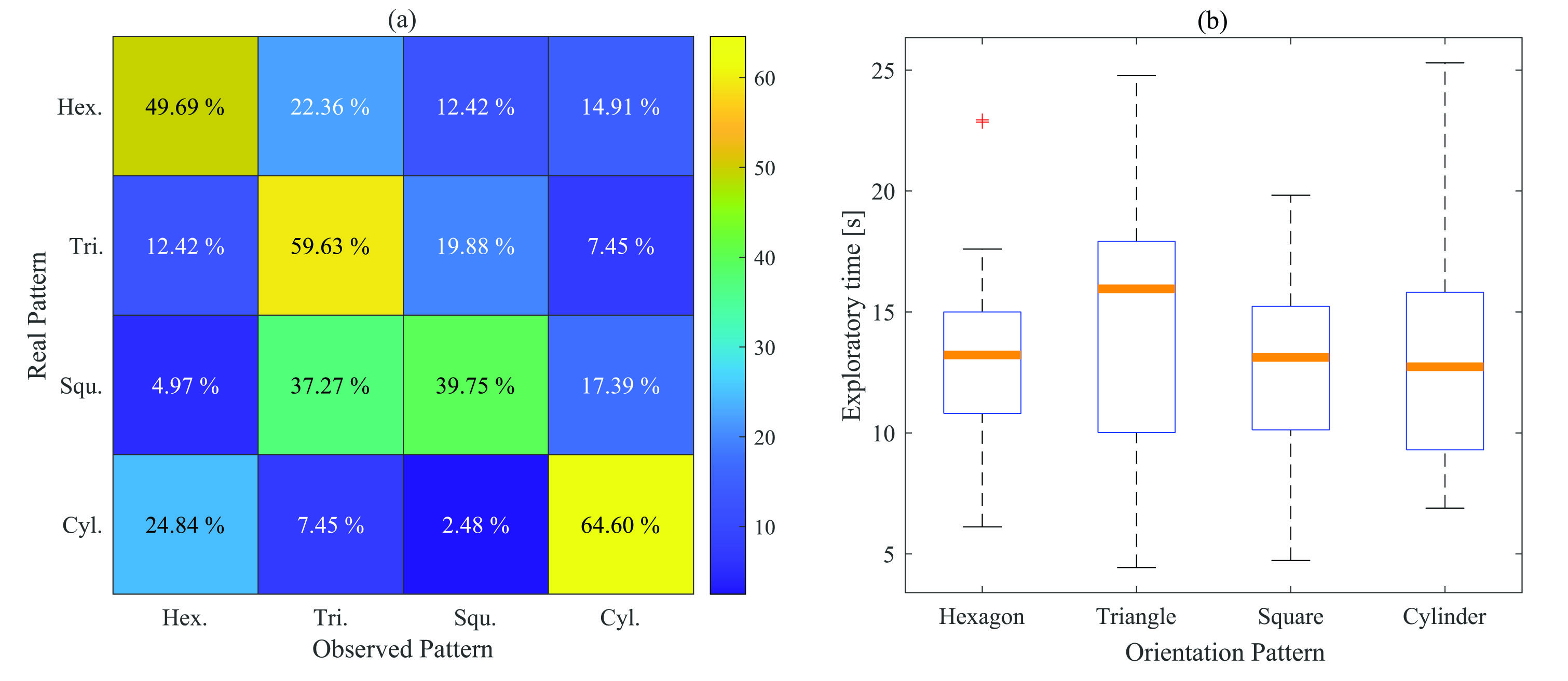}
\end{center}
\caption{Experiment 2. (a) Confusion Matrix about 3D Shape recognition rate. (b) Exploratory time comparative evaluation for 3D shape recognition}\label{bs3}
\end{figure*}

 Fig.\ref{bs3} (a) shows a numerical comparison of the effective shape recognition level for each proposed pattern. We observe that the four patterns have a different range of recognition, being the \textit{square} pattern pointed the lowest (40\% accuracy) and the \textit{cylinder} pattern the highest (65\% accuracy). The result also indicates that the \textit{square} pattern is frequently confused with the \textit{triangle} pattern (37\% error),  and the \textit{cylinder} is confused with \textit{hexagonal} pattern (25\% error). 

The experiment also includes an analysis of exploration time. Fig.\ref{bs3} (b) shows that the recognition time for the majority of the volunteers ranges between 8 and 18 \textit{seconds}. The median time is close to 13 \textit{seconds} also for all of the cases, except for the \textit{triangle} pattern which median exploratory time is 16 \textit{seconds}.

\section{Conclusion}
In this paper, we mainly developed a haptic feedback component of the virtual reality system for remote training. We implemented a simple tactile communication capable of transmitting shape sensations produced at the moment of manipulating 3D objects with two fingers: thumb and index fingers. The follower side comprises a tactile-sensor glove and the leader side comprises an electro-tactile display glove.

We tested our system with two experiments: static shape perception and dynamic pattern perception, both assuming the situation of grasping a bar-like object. The results confirmed our expectations, that this system has the ability to deliver information of 3D bar-like object.

There are several limitations to the current work. The visual part of the system is incomplete; the follower side should see the hand gesture of the trainer, and the leader side should see 3D visual information of the follower by the use of 3D display technologies. Tactile display and sensor are slightly small, and it must be enlarged to cover the whole fingertips. Roughness and temperature sensations must be considered for providing material sense. All these will be handled in our future work.

%
%
%
%


\begin{thebibliography}{}




\bibitem{Susumu} 

Tachi, S.: Tele-existence - Toward Virtual Existence in Real and/or Virtual Worlds, In. Proc. ICAT '91, pp.85-94, 1991

\bibitem{Kim kang} 
Kim, S.-C., Kim, C.-H., Yang, G.-H., Yang, T.-H., Han, B.-K., Kang, S.-C., Kwon, D.-S.: Small and lightweight tactile display (SaLT) and its application. In: Proceedings WorldHaptics, pp. 69–74 (2009)


\bibitem{Beebe} 
D. Beebe, D. Denton, R. Radwin, and J. Webster, “A silicon-based tactile sensor for finger-mounted applications,” IEEE Trans. Biomed. Eng., vol. 45, pp. 151–159, 1998



\bibitem{Choi} 
Choi, I., Hawkes, E.W., Christensen, D.L., Ploch, C.J., Follmer, S.: Wolverine: a wearable haptic interface for grasping in virtual reality. In. Proc. IROS, pp. 986-993.



\bibitem{Sarakoglou} 
Sarakoglou, I., Tsagarakis, N., Caldwell, D.G.: A portable fingertip tactile feedback array – transmission system reliability and modelling. In. Proc. WHC2005, pp. 547-548


\bibitem{Saunders} 
F.A. Saunders, In Functional Electrical Stimulation: Applications in Neural Prostheses, ed. by F.T. Hambrecht, J.B. Reswick (Marcel Dekker, New York, 1977), pp. 303–309



\bibitem{P.Bach} 
P. Bach-y-Rita, K.A. Kaczmarek, M.E. Tyler, J. Garcia-Lara, Form perception with a 49-point electrotactile stimulus array on the tongue. J. Rehab. Res. Dev. 35, 427–430 (1998)



\bibitem{Kaji electro} 
Kajimoto H. (2016) Electro-tactile Display: Principle and Hardware. In: Kajimoto H., Saga S., Konyo M. (eds) Pervasive Haptics. Springer, Tokyo


\bibitem{Yem Kaji} 
Yem V., Kajimoto H., Sato K., Yoshihara H. (2019) A System of Tactile Transmission on the Fingertips with Electrical-Thermal and Vibration Stimulation. In. Proc. HCII 2019



\end{thebibliography}

\end{document}